\documentclass[11pt,twoside]{article}
\usepackage{cozumel2005}
\usepackage{epsf}
\usepackage{lscape}
\begin{document}
\title{Galactic Open Clusters}
\author{Ted von Hippel}
\affil{University of Texas at Austin}

\begin{abstract} 
The study of open clusters has a classic feel to it since the subject
predates anyone alive today.  Despite the age of this topic, I show via an
ADS search that its relevance and importance in astronomy has grown faster
in the last few decades than astronomy in general.  This is surely due
to both technical reasons and the interconnection of the field of stellar
evolution to many branches of astronomy.  In this review, I outline what
we know today about open clusters and what they have taught us about a
range of topics from stellar evolution to Galactic structure to stellar
disk dissipation timescales.  I argue that the most important astrophysics
we have learned from open clusters is stellar evolution and that its most
important product has been reasonably precise stellar ages.  I discuss
where open cluster research is likely to go in the next few years, as
well as in the era of 20m telescopes, SIM, and GAIA.  Age will continue
to be of wide relevance in astronomy, from cosmology to planet formation
timescales, and with distance errors soon no longer a problem, improved
ages will be critically important to many of the most fascinating
astrophysical questions.  
\end{abstract}

\section{Minimal History of Galactic Open Cluster Research}

Rather than linearly review what has been learned about open clusters
to date or update the excellent review of Friel (1995), I will use a
44 year old paper by Sandage (1961) as a departure point to cast the
problems and opportunities presented by open clusters in perspective.
From that brief introduction, I will take a look at 1) the general
properties of Galactic open clusters, 2) the most important science
to come from open cluster research, and 3) the diverse range of other
science derived from open cluster research.  All of these topics will
be highlighted with work from only the last 1.5 years, since these
research fields are so active.  I conclude with a brief discussion
of the opportunities and challenges of the next ten years.

In 1961, Sandage wrote a short paper entitled ``The ages of the Open
Cluster NGC 188 and the Globular Clusters M3, M5, and M13 compared
with the Hubble Time.''  Based on his observations and models by Hoyle
(1959), Sandage reported that ``the ages of the clusters are computed to
be 16 x 10$^9$ years for NGC 188, ...''.  From these age determinations,
and an assumption ``if H=75 km/sec/10$^6$pc'', then ``available data
are therefore inconsistent.  Changes in (1) the stellar evolution
time scale, (2) the value of the Hubble constant, (3) the observational
redshift-magnitude data, or (4) cosmological theory seem to be required
at this point.''  In fact, item 1 changed by a factor of two and item
4 has become considerably more complicated than the $\Omega_{matter}$ =
1 Universe many imagined in 1961.  For at least 40 years, observations
of open clusters and their theoretical interpretation has been deeply
coupled to our understanding of stellar evolution, the age of the
Universe, and cosmology.  I will argue below that the comparison of open
clusters to stellar evolution models and the derivation of stellar ages
have been the most important science to come of open cluster studies.
These studies alone are only part of the vast literature about or relying
on open clusters.  A statistical look at the literature will help put
open cluster research into perspective.

Figure 1 approximates the growth in the research on open and globular
clusters over the last 45 years by reporting the number of papers
listed by ADS with the phrase ``open cluster'' or ``globular cluster''
in the abstract.  Such papers are either studies of these star clusters,
or are dependent on such studies.  The line in the figure is the growth
rate in the astrophysics literature, as determined by Abt (1998),
normalized to the number of open cluster papers in 1960.  Open cluster
research followed the general growth trend in astrophysics from 1960
through the late 1980's, after which the expansion of open cluster
research was more pronounced.  Thus, despite being a classical subject in
astronomy, open cluster research remains vital and increasingly relevant.
I interpret the current growth in open cluster research to be the result
of a renewed desire for precision in stellar population studies plus
the steady growth in CCD mosaic sizes.  In contrast, globular cluster
research appears to have expanded rapidly in the mid-1970's, followed
the general growth trend, then leveled off during the last ten years.

\begin{figure}[!ht]
\plotfiddle{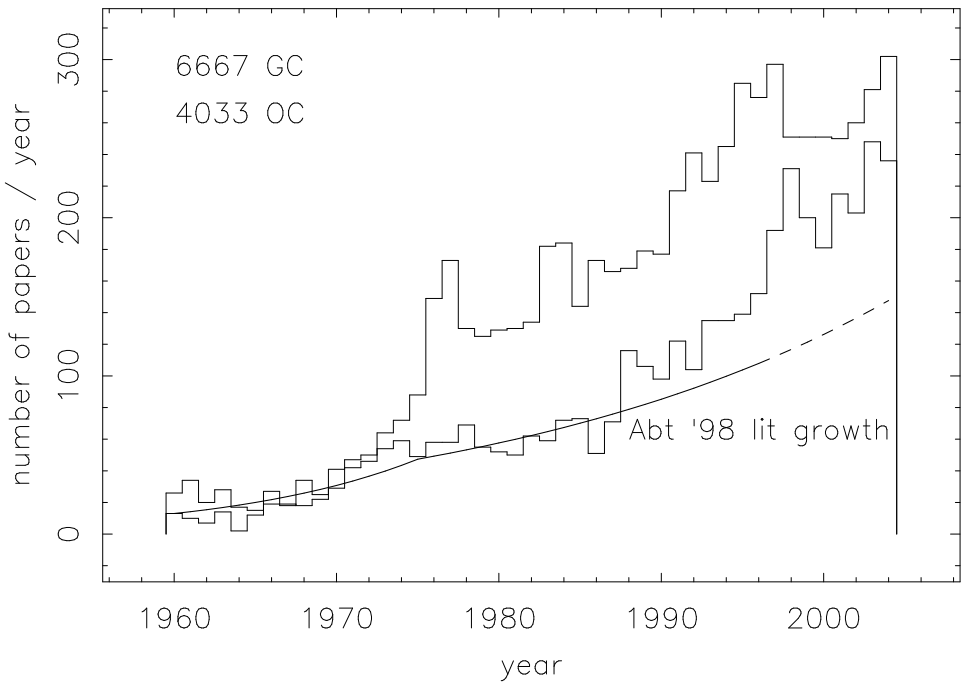}{3.7in}{270}{55}{55}{-200}{320}
\caption{The growth of astronomical literature as a function of year.
The overall growth as given by Abt (1998) is indicated by the solid
line and extrapolated with the dashed line.  The number of papers found by
ADS with the phrases ``open cluster'' and ``globular cluster'' are given
by the lower and upper histogram, respectively, with the number of papers
in each category listed at the upper left of the figure.}
\end{figure}

\section{Overview of Galactic Open Cluster Properties}

Galactic open clusters have typical distances of one to a few kpc, largely
due to observational bias.  They are found primarily in the Galactic
plane (Fig.\ 2, see Jean-Claude Mermilliod's excellent website,
http://obswww.unige.ch/webda/ for the data that went into this and some
of the following plots), since many of these objects are young (Fig.\ 3).
A more sophisticated study of the age distribution of open clusters is
given by Salaris, Weiss, \& Percival (2004).

\begin{figure}[!ht]
\plotfiddle{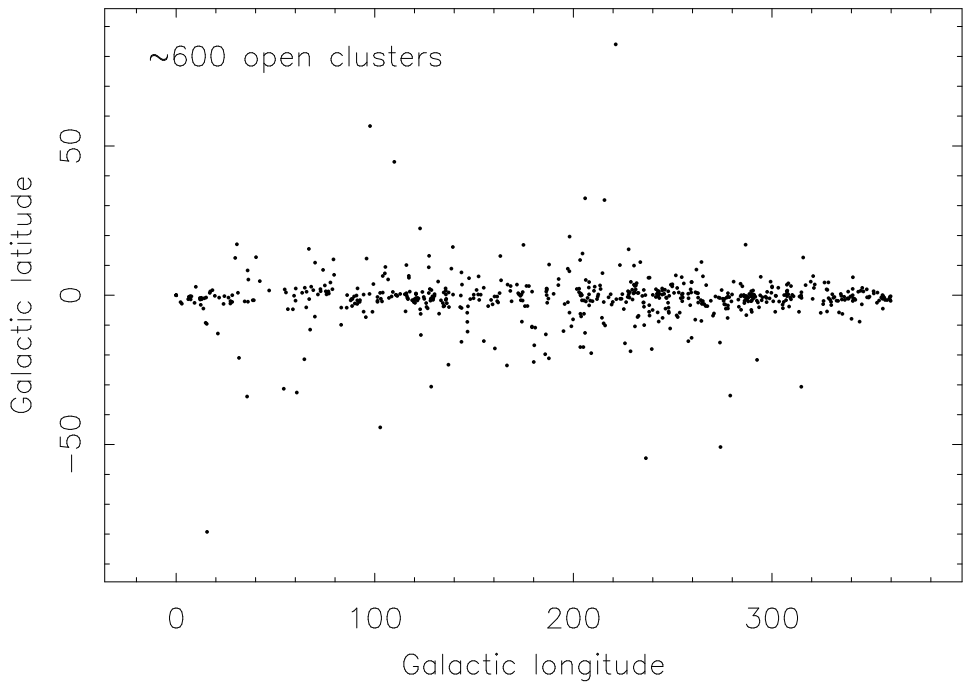}{3.7in}{270}{55}{55}{-200}{320}
\caption{The Galactic longitude and latitude location of all open clusters
listed by J.-C. Mermilliod in his web database (obswww.unige.ch/webda/).}
\end{figure}

\begin{figure}[!ht]
\plotfiddle{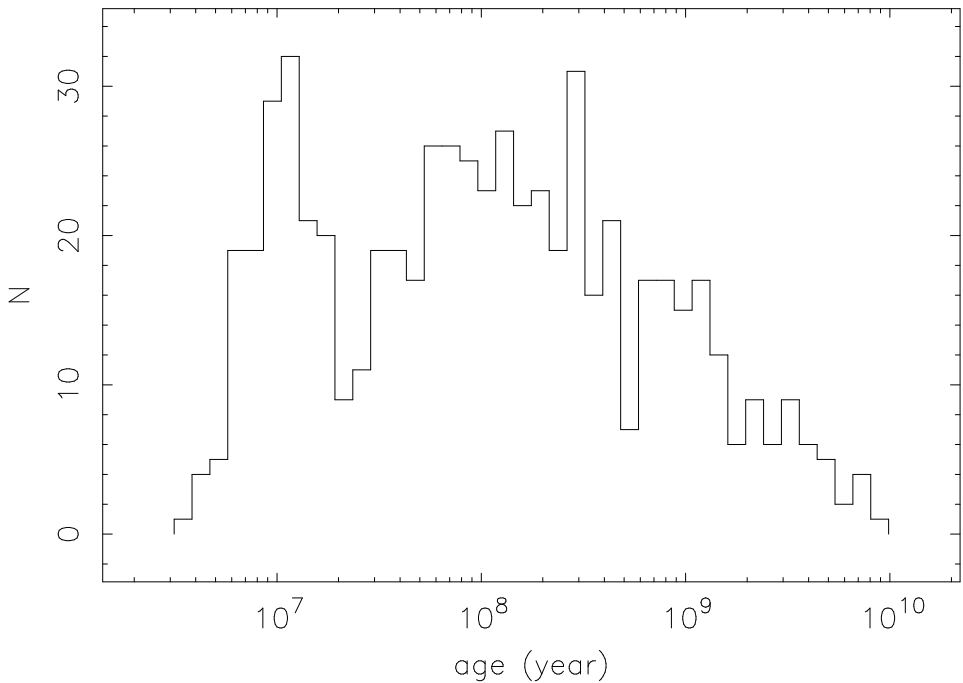}{3.7in}{270}{55}{55}{-200}{320}
\caption{The Galactic age distribution of all open clusters with an
estimated age value listed by J.-C. Mermilliod in his web database.}
\end{figure}

Open clusters show no obvious age-metallicity relation (e.g., Janes 1979;
Friel 1995), but they do show a metallicity gradient as a function of
Galactocentric distance (e.g., Salaris et al.\ 2004).  Careful studies
of open clusters, planetary nebulae, and B stars by a number of groups
indicate that $\sim$ 8 Gyr ago, the Galactocentric metallicity gradient
was $\sim$ $-0.1$ dex kpc$^{-1}$, that it has flattened with time, and
that it is now of order $-0.04$ dex kpc$^{-1}$ (Daflon \& Cunha 2004).
This result shows the timescale of enriching and mixing within the
Galactic disk.

The age-metallicity distribution for open and globular clusters (Fig.\ 4)
shows a complete lack of metal-poor young clusters and shows a tantalizing
gap between the ages and metallicities of open clusters versus globular
clusters.  I interpret that gap not as a fundamental statement of star
formation efficiency at [Fe/H] = $-0.6$ to $-0.8$, but rather as evidence
that the Galaxy evolved rapidly through this intermediate metallicity
and/or the star clusters formed at this metallicity survived in even lower
fractions than halo globular clusters, only 1\% of which survive to date,
or open clusters, which survive on Gyr timescales in even lower fractions.

\begin{figure}[!ht]
\plotfiddle{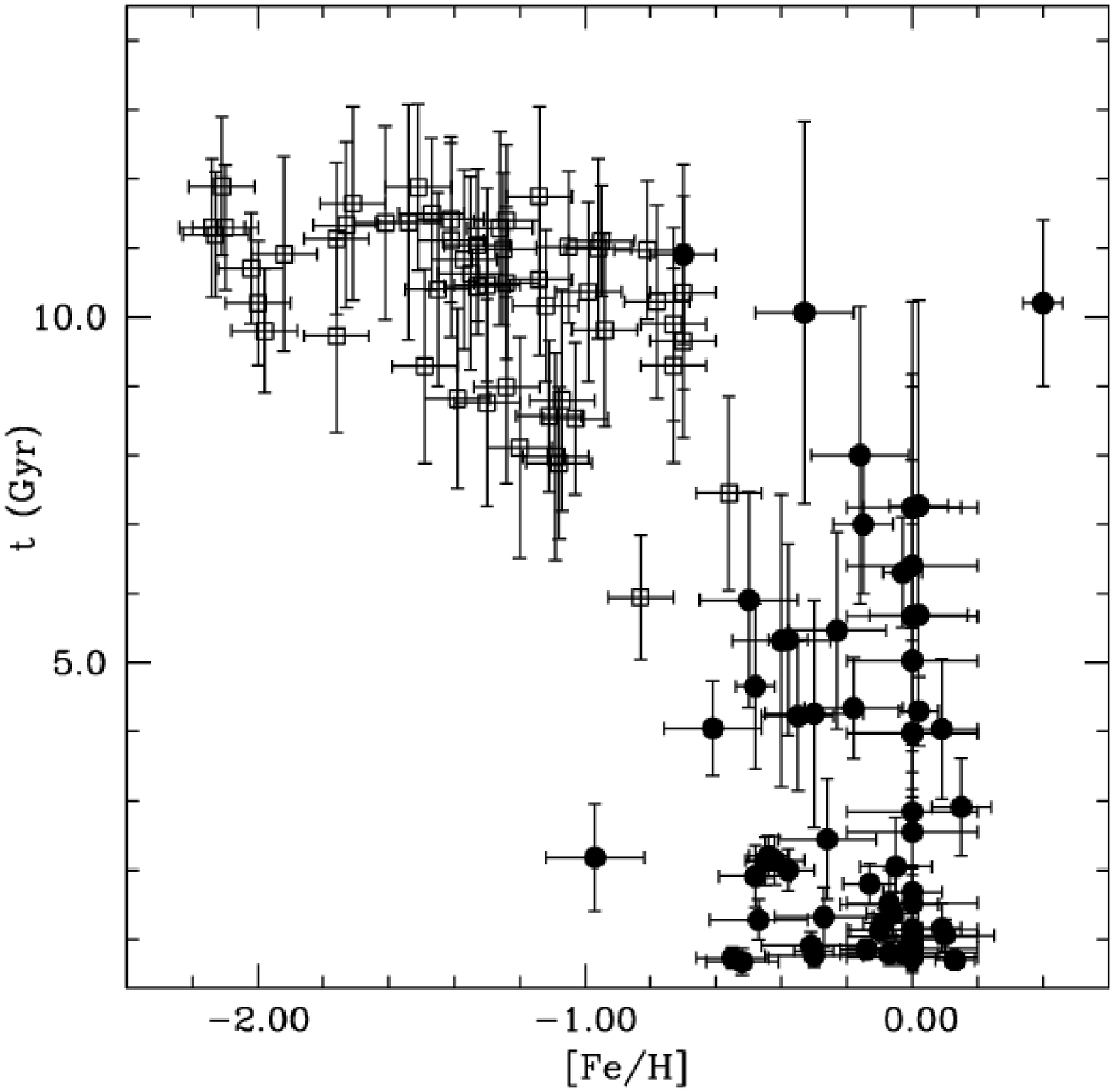}{4.1in}{0}{55}{55}{-180}{-75}
\caption{The age vs.\ metallicity distribution for a large number
of open and globular clusters (from Salaris et al.\ 2004, fig.\ 10).
The open and globular clusters are indicated by the filled circles and
open squares, respectively.} 
\end{figure} 

Interestingly, despite the very low dispersions of open clusters,
typically $\leq$ 0.5 km s$^{-1}$ (Mathieu 1985; Zhao \& Chen 1994),
open clusters lose members steadily, and most eventually dissolve into
the Galactic field star population.  There is a continuum between open
clusters and moving groups and between moving groups and field stars.
And the smallest remnants of dissolved open clusters, binary stars,
can be thought of as the simplest open clusters.  For non-contact
binaries, complicated N-body dynamics reduce to well understood 2-body
interactions, and these objects have been important for stellar mass
studies and distance estimates for about a century.

Since open clusters dissolve readily, the oldest open clusters place
only a lower limit on the age of the Galactic disk.  Still, that limit
is both important and reasonable, with NGC 6791 and Be 17 being $\sim$10
Gyr (Salaris et al.\ 2004).  It is possible, though not convincingly
demonstrated, that some of the globular clusters (e.g., 47 Tuc) belong
to the thick disk.  If this is correct, then these objects place the
thick disk at $\sim$12 Gyr (Liu \& Chaboyer 2000) old, at or just younger
than the age of the Galactic halo and apparently substantially older
than the thin disk.

Our discussion of the properties of open clusters has unavoidably touched
on properties of the Galaxy.  While important, in my opinion the most
important result from open cluster research is the steady and detailed
testing and refining of stellar evolution models.

\section{The Most Important Concepts Open Clusters Have Taught Us}

We take it as a given that we can model stars and their evolution and
that we know the age of any stellar population that we have studied in
sufficient detail.  But this capability of ours should not be treated as
an obvious and trivial extension of physics learned in labs on Earth.
Certainly understanding stellar evolution is an extension of physical
processes learned in our labs.  But understanding stellar evolution is
a lot more than that -- it is an amazing success, one of the greatest
scientific triumphs of the last 50 years.  That these distant objects
are understandable, even predictable, is a testament to the general
applicability of scientific principles.  And the derived properties of
cluster stars underpins much of Galactic and extragalactic astronomy.
In my opinion, the most important parameter to come from the mature field
of stellar evolution is the age of stars and stellar systems.  Just as
the ability to date geological strata rapidly advanced geology, and just
as molecular clocks are a key to unraveling evolutionary history of life
on Earth, so too the ability to age date stars drives our cosmology.
Additionally, the difference between ages often yields important
astrophysical timescales.  For instance, the difference between the age of
the Sun (e.g., Gough 2001) and the planets (e.g., All\`egre, Manh\`es, \&
G\"opel 1995) constrains the timescale for planet formation.  Likewise,
the difference between the ages of the oldest stars in the Galaxy and
the age of the Universe provides the timescale for galaxy formation.

How, exactly, do we determine stellar ages of open clusters?  The most
common tool, and probably the second most common diagram in astronomy
after the spectrum, is the color-magnitude diagram (CMD).  The location
of stars in the CMD will provide a model-dependent set of correlated
constraints on the cluster's age, metallicity, distance, and reddening.
Yet, often times cluster CMDs are contaminated by foreground and
background Galactic field stars.  Such contaminants can be removed by
proper motion (e.g., Platais et al.\ 2003) or radial velocity cuts (e.g.,
Daniel et al.\ 1994), or statistically via comparison with an adjacent
field or even with Galactic star count models.  Even with the addition
of outside information to remove contaminating field stars, and despite
the maturity of the field of stellar evolution, deriving stellar ages
from CMDs is fraught with difficulty.  There are uncertainties in theory
(see Cassisi, this volume) and uncertainties in the transformation of
theory to observations (see VandenBerg, this volume).  For example, the
detailed studies of the old ($\sim$4 and $\sim$6 Gyr, respectively) open
clusters M67 and NGC 188 by VandenBerg \& Stetson (2004), particularly
their figures 3 and 7, show how cluster CMDs can be fit with numerous
combinations of cluster age, [Fe/H], stellar helium content, and distance.
Additionally, simultaneous fits to CMDs in different colors such as
$B-V$ and $V-I$ often do not match the location of the main sequence,
turn-off, subgaint branch, or red giant branch in all the CMDs (e.g.,
for NGC 6791 see Chaboyer, Green, \& Liebert 1999 and for NGC 188 see
Sarajedini et al.\ 1999).  The theory of stellar evolution is mature
and the quality of the data we obtain can be very high, yet important
theoretical details still need to be fully understood and the derived
parameters, in particular age, are not as precise as we wish them to be.
Typical age uncertainties, even in the most carefully studied clusters,
are $\pm$ 20\%.

Open clusters have not been mined to their full potential, however,
and even without improving our telescopes, stellar evolution and
stellar atmospheres can be and are being improved and tested on new
observations.  Model ingredients can be tested on stars in particular
states of evolution, especially when there is abundant information
on the stellar properties, such as $T_{eff}$ and log(g) from stellar
atmospheres, radii from angular measurements or the fortuitous case of
eclipsing spectroscopic binaries, and masses from binaries or potentially
gravitational redshifts (von Hippel 1996).  Another basic model ingredient,
the helium content of stars, cannot currently be derived from stellar
evolution models, as it is the uncertain mass loss process that drives
this quantity, not the amount of helium created by the previous generation
of evolving stars.  Yet this number too is approachable via observations
of open clusters, particularly where multi-color CMDs and independent
radii for low mass cluster stars can be obtained.

\section{Other Important Concepts Learned from Open Clusters}

Open clusters, each containing stars with a range of masses but only a
single age, abundance pattern, and distance, have provided tremendous
insight into a range of astrophysical problems, from those intimately
related to stellar interiors, atmospheres, and evolution, to problems
of cosmology, to disk and planet formation.

Figure 5, for instance, shows the lithium abundance vs.\ $T_{\rm eff}$
for NGC 2547 members and the expectations from 30 Myr and 50 Myr models,
appropriate for the cluster age (Jeffries \& Oliveira, 2005).  The degree
of Li depletion as a function of stellar effective temperature and age
has been the subject of many studies since the discovery by Boesgaard \&
Tripicco (1986) of Li depletion in Hyades F stars, and importantly teaches
us about stellar surface mixing as well as the photon to baryon density
during Big Bang nucleosynthesis (Richard, Michaud, \& Richer 2005).
Furthermore, placing Li age constraints back into the HR diagram or CMD
(see Fig.\ 6 and also the contribution by E. Mart\'in, this volume),
brings the theory of pre-main sequence or main sequence stellar evolution
into direct conflict with theories of stellar surface convection and
the destruction of Li via nuclear processes.  Such a confrontation,
here based on the derived ages that best match each theory to the data,
could produce the same derived age, or could produce different ages.
Coincidences of age are unlikely if either or both theories have
substantial problems, especially after studying multiple clusters.

\begin{figure}[!ht]
\plotfiddle{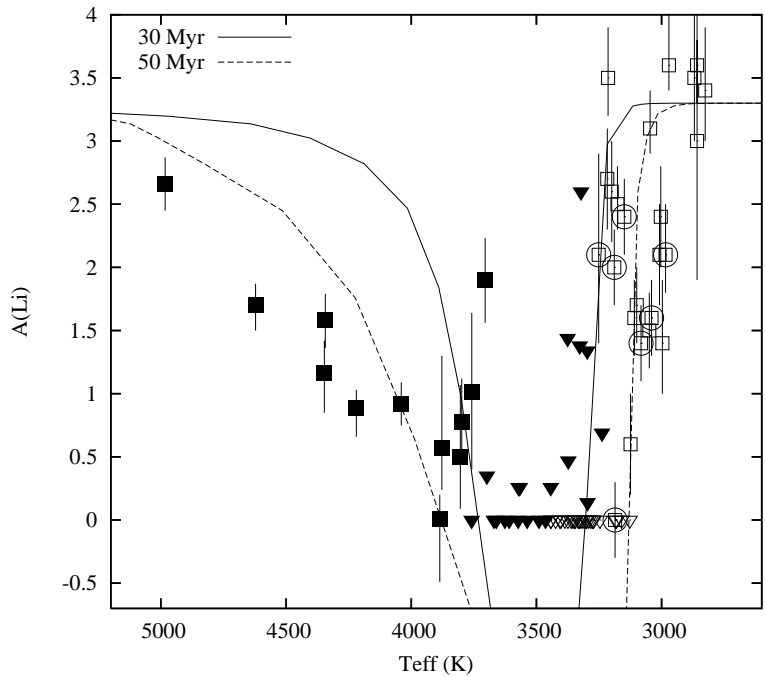}{4.0in}{0}{60}{60}{-180}{0}
\caption{Lithium abundances vs.\ $T_{\rm eff}$ for NGC 2547 members
showing Li depletion as a function of $T_{\rm eff}$ compared to
expectations for 30 Myr and 50 Myr models (from Jeffries \& Oliveira 2005,
fig.\ 4).} 
\end{figure} 

\begin{figure}[!ht]
\plotfiddle{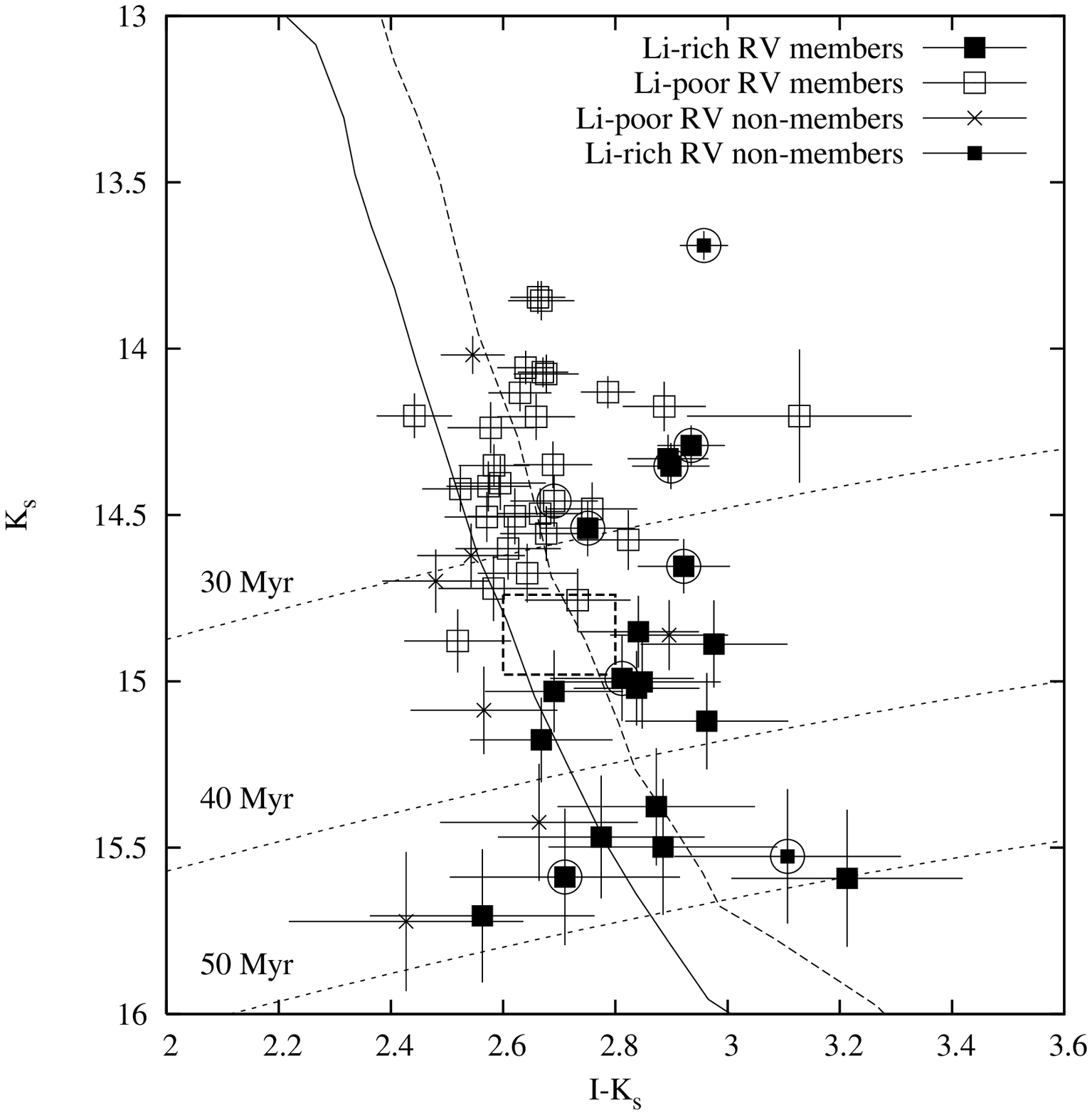}{4.1in}{0}{60}{60}{-180}{-25}
\caption{$K$ vs.\ $I-K$ CMD for NGC 2547 showing standard pre-main
sequence isochrones compared with stellar photometry, as well as
the time-dependent Li depletion boundary (from Jeffries \& Oliveira 2005,
fig.\ 7).} 
\end{figure} 

Skipping to another timescale of stars, tidal effects should circularize
orbits in non-contact, but closely orbiting, binaries.  The distance at
which stellar orbits have become circularized should therefore increase as
a function of time, and this has been carefully studied both to understand
stellar structure and to create a new stellar chronometer that might be
useful for field star studies (e.g., Mathieu, Meibom, \& Dolan; Meibom \&
Mathieu 2005).  Tidal circularization is a particular form of angular
momentum evolution, and other forms, for instance braking in stellar
rotation (e.g., Queloz et al.\ 1998; Melo, Pasquini, \& De Medeiros 2001)
and star-disk coupling (Carpenter et al.\ 2005) can also be productively
studied in open clusters.

Another important topic of stellar evolution, and one also coupled
to stellar ages, is the initial-final mass relation.  This is the
relationship, often assumed to be monotonic and constant from one
cluster to another, between the zero-age main sequence mass of a star,
and the mass of its white dwarf (WD) descendant.  This relationship is the
empirical outcome of the poorly understood process of mass loss during
late stages of stellar evolution.  It may turn out that mass loss is
dependent on stellar rotation, binarity, metallicity, or magnetic fields,
and so perhaps no single initial-final mass relation is valid.  On the
other hand, wild differences in this relationship from cluster to cluster
have not been found to date.  The upper mass limit for the initial-final
mass relation, i.e., the highest mass star that will evolve into a WD,
is also an important, and as yet imprecisely constrained number, most
likely between 7 and 9 $M_{\sun}$.  This number, in turn, is an important
constraint for the theory of stellar evolution and an ingredient in
stellar population chemical evolution models.  The initial-final mass
relation also depends on models of main sequence stellar evolution
to determine cluster ages and thereby masses of progenitors, and it
furthermore is required to use field WDs as chronometers, as their cooling
times need to be added to their progenitor lifetimes in order to derive
their ages.  For excellent recent studies of the initial-final mass
ratio see Williams, Bolte, \& Koester (2004) and Kalirai et al.\ (2005).

A relatively new topic for open cluster research is the connection of
open cluster ages to stellar IR excesses.  Since we cannot reliable
date most single field stars, young clusters with known ages provide
the only way to study the evolution of disks.  Figure 7, from Mamajek
et al.\ (2004), displays the N-band excess, a proxy for disk mass, for a
variety of young clusters.  Disk dissipation takes on the order of $10^7$
years, which is an important constraint for theories of planet formation.
This topic will be of increasing importance as star, disk, and planet
formation models and data advance with missions such as Kepler, TPF,
and Darwin.

\begin{figure}[!ht]
\plotfiddle{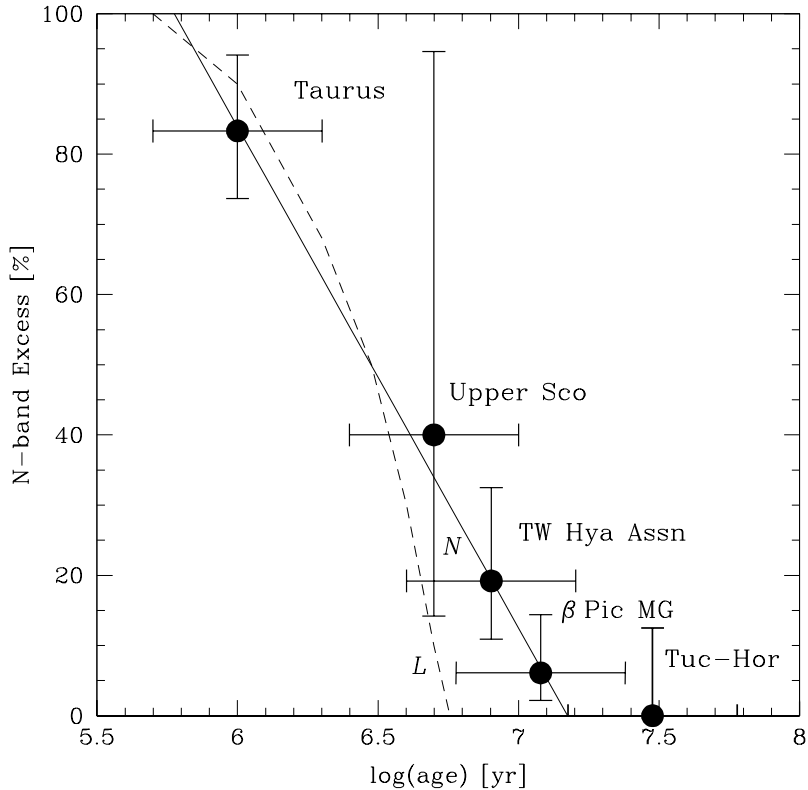}{4.1in}{0}{55}{55}{-180}{-75}
\caption{$N$-band excess, which is a proxy for disks, for young clusters
as a function of cluster age (from Mamajek et al.\ 2004, fig.\ 3).} 
\end{figure} 

A classic use of open clusters is to calibrate the distance ladder.
Every open cluster is amenable to main sequence fitting to derive
distance, especially once the cluster metallicity is determined.
This is most useful if the reddening is low or independently measured,
and there are many such open clusters.  In addition, some open clusters
contain Cepheids (e.g., Hoyle, Shanks, \& Tanvir 2003), others contain
eclipsing binaries (e.g., Southworth, Maxted, \& Smalley 2005), and a
few open clusters are rich enough to calibrate the red clump magnitude
(e.g., Grocholski \& Sarajedini 2002).

No discussion of open clusters would be complete without mentioning the
Initial Mass Function (IMF), yet I will only discuss it briefly, in part
because numerous reviews have been dedicated to the subject (Miller \&
Scalo 1979; Scalo 1986; Mould 1996; Chabrier 2003), and even entire
conferences (The Stellar Initial Mass Function: 38th Herstmonceux
Conference in 1998), and because the IMF and its derivation are
discussed by others at this conference (see contributions by P. Kroupa
and J. Ma\'iz Apell\'aniz).  Let me merely point out that the observed
mass distribution of stars in open clusters results from a combination
of the IMF, stellar evolution, and the cluster's dynamical evolution,
and these complexities have, to date, made it difficult to discern
whether there is a universal IMF, or whether it varies meaningfully from
cluster to cluster.  My feeling is that the quality of the data are now
good enough to start to discern significant differences from cluster
to cluster.  Whether these differences will be meaningful in the context
of Galactic chemical evolution remains to be seen, and the evidence does
not support the range of IMFs often used in extragalactic studies to
reconcile necessarily sparse observations with stellar population theory.
One of the most impressive aspects of the IMF is that in some clusters,
e.g., Lambda Orionis (Barrado y Navascu\'es et al.\  2004), observations
have now shown that the mass function is smooth until well below
the hydrogen burning limit.  In this cluster, objects are observed
down to $\sim0.3 M_{\sun}$.  Even though this particular mass value
is highly uncertain due to difficulties in modeling these stars,
a smooth IMF into the brown dwarf regime is a fundamental statement
about the star formation process and the limitations of simple Jeans
mass explanations for star formation.  Clearly more complicated physics
are involved in star formation than simply gravity and thermal energy.
Turbulence, fragmentation, and interaction most likely play roles in
some complicated and probably subtle manner.  Additionally, the fact
that we can now readily observe brown dwarfs in open clusters provides
important opportunities to calibrate their mass-luminosity relationship
and to study their atmospheres, formation, and evolution.

\section{Where to Next?}

\subsection{Technology}

The current generation of 8-10m telescopes are just now being applied to
open cluster problems.  Their multiplexing spectrographs are excellent
machines to follow-up on wide-field studies at smaller telescopes.
Examples of this work are white dwarf spectroscopy to $V \approx 23$
(Kalirai et al.\ 2005) using both Keck and Gemini.  In just over a
decade, we should have 20m class telescopes available, and these will
make it substantially easier to study brown dwarfs and disks in nearby
open clusters.  On about the same timescale, NASA's Space
Interferometry Mission, will derive exquisite parallax distances ($\sim5$
microarcseconds, Chaboyer et al.\ 2005) to important globular clusters.
The GAIA mission, a cornerstone mission for ESA, will obtain similar
quality astrometry for $\sim10^9$ stars, including all stars brighter
than $V \approx 20$ (de Zeeuw 2005).  Distance, which is often the single
greatest source of uncertainty in open cluster research, will become a
precision parameter.  Stellar evolution theory will no longer be able
to hide many of its more subtle inadequacies.  To take full advantage
of these exquisite ages we need to push abundance errors down from their
current level of 0.1-0.2 dex, to $\leq$ 0.05 dex.  This should be possible
with careful work and high quality spectroscopy.

\subsection{Further Refine Theory}

Improved distances and metallicities will drive refinements in theory.
Yet even before SIM and GAIA, we can hope for improvements as there are
a number of approaches that have not been fully exploited.  The detailed
shape of isochrones and their fits to cluster CMDs as well as stellar
number counts along the isochrones for open clusters covering a range
of ages have already been used to tune stellar evolution parameters
such as overshoot (e.g., Demarque et al.\ 2004).  This process can
be continued for other clusters and with better data, and other stellar
evolution parameters can be improved upon via this technique.  The overly
simplistic mixing length theory is one example that needs refinement,
and this is being pursued (Canuto, Goldman, \& Mazzitelli 1996; Kupka \&
Montgomery 2002).  The spectroscopy of highly evolved stars also yields
clues to dredge-up and other mixing processes, and the study of pulsating
stars holds out hope for refining stellar structure in a manner entirely
independent of matching the external properties of stars to interior
models plus atmosphere models.

Another technique where observations can test and help guide theory,
is the simultaneous age dating of cluster WDs and cluster main sequence
turn-off (MSTO) stars.  Since a single cluster has one age, both the WD
cooling ages and the main sequence stellar evolution ages should agree.
The WD observations have been demanding in the past, but with the current
abundance of 8-10m telescopes as well as HST with the ACS instrument,
more of the very faint (=old) WDs in open clusters have been observed.
Open clusters with a range of ages and abundances are now being studied
via this technique (see von Hippel 2005, and references therein) and
one globular cluster, M4, has been observed to sufficient depth for a
WD age (Hansen 2004).  In this volume, Jeffery et al.\ demonstrate a
Bayesian modeling technique that holds promise for deriving cluster WD
ages even in cases where the faintest WDs are too faint to be observed.
Their technique, while still being developed, should allow us to add more
distant star clusters to our studies and widen the applicability of the
WD technique.  Figure 8 presents the current status of cluster age studies
where both the WD cooling technique and the traditional main sequence
evolutionary age technique have been applied.  The good news is that
stellar evolutionary timescales appear firm to 2 Gyr and probably 4 Gyr.
(I emphasize the match to 2 Gyr and the uncertainty at 4 Gyr since the
4 Gyr cluster, M67, anchors the upper age end by itself and since the
WD age was derived after a statistical background subtraction (Richer
et al.\ 1998)).  Many more clusters can be added to this comparison,
and this should and will be done.

\begin{figure}[!ht]
\plotfiddle{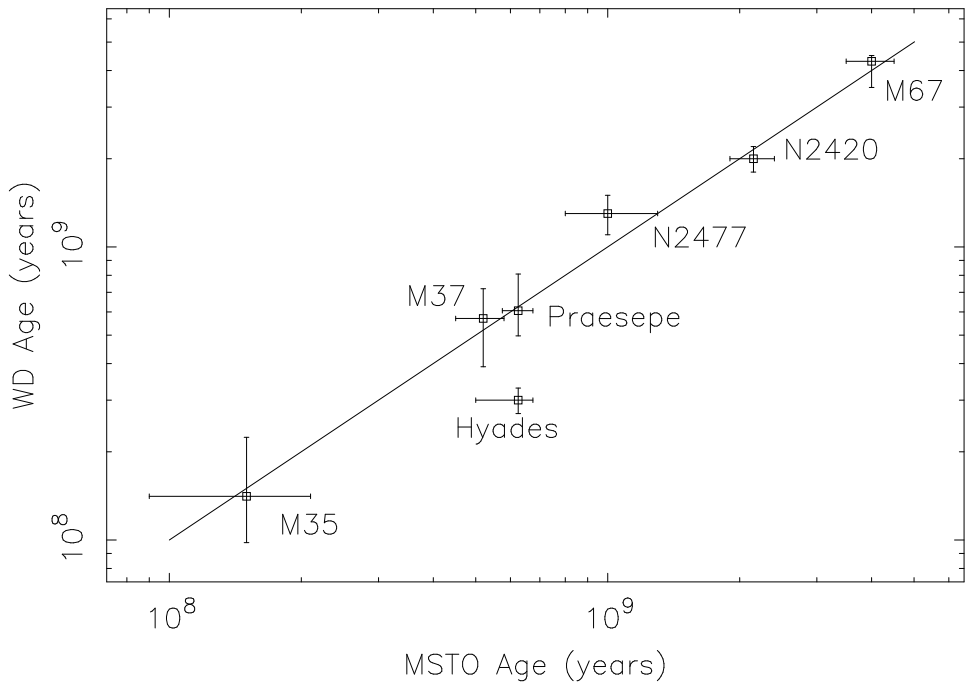}{3.7in}{270}{55}{55}{-200}{320}
\caption{Main sequence turn-off age vs.\ white dwarf age for seven
clusters (from von Hippel 2005, fig.\ 1).} 
\end{figure}

It is important to test the most important parameter of our stellar
evolutionary models whenever possible.  At this conference, during
Eduardo Mart\'in's talk, I realized there is hope for an WD age vs.\
MSTO age vs.\ Li age test for at least a few young clusters.  And there
may be other multiple age tests, at least at the level of a consistency
check, for instance with stellar circularization or disk dissipation.
The goal in these tests is to bring into potential conflict the most
important derived quantities of the theory, and thereby both test the
theory and provide an empirical estimate of the accuracy of the predicted
quantities.

\subsection{Highest Impact Science}

I believe that stellar ages will remain the highest impact science to
come from open cluster research for some time.  Refinements in theory
and improvements in distances, metallicities, etc.\ can drive age
precisions from the present best case of $\sim20$\% to perhaps 5\%.
Stellar ages, in turn, will refine our understanding of the formation
sequence and structure formation process in the Galaxy.  A refined stellar
evolution theory coupled with exquisite GAIA distances to field stars
will also make it possible to derive ages for many of the slightly to
moderately evolved field stars.  Improved age precision will in turn
be necessary for answering questions in new fields, such as stellar
disk dissipation and planet formation timescales.  It is also possible
that our colleagues will soon discover planets in open clusters, via the
transit, radial velocity, or direct imaging techniques.  The properties
of planets in systems with known ages will be substantially more useful
for understanding planet formation and evolution than similar planets
found around field stars of uncertain age.

\section{Conclusion}

The study of open clusters has a classic feel to it since the subject
predates any of us.  Despite the age of this topic, its relevance and
importance in astronomy has grown faster in the last few decades than
our field in general.  This is due to both technical reasons and the
interconnection of the field of stellar evolution to many branches
of astrophysics.  Large field of view imagers on 4m class telescopes,
multi-plexing spectroscopy on 8-10m telescopes, and HST are all
contributing to the rapid growth in open cluster research.  The topics
open clusters can address, ranging from subtleties of stellar evolution to
the distance ladder to disk dissipation and planet formation timescales,
indicates that these star clusters that we have been fortunate to live
among will continue to be important research arenas for decades to come.

\acknowledgements
I would like to thank David Valls-Gabaud and the SOC for inviting
me to present this review, and the LOC for their organizational,
professional work.  This material is based upon research supported by the
National Aeronautics and Space Administration under Grant No.\ NAG5-13070
issued through the Office of Space Science, and by the National Science
Foundation through Grant AST-0307315.


\end{document}